\documentclass[fleqn,usenatbib]{mnras}

\usepackage{newtxtext,newtxmath}

\usepackage[T1]{fontenc}
\usepackage{ae,aecompl}

\usepackage{graphicx}	
\usepackage{amsmath}	
\usepackage[usestackEOL]{stackengine}

\title[Multi-messenger LISA inference]{Multi-messenger parameter inference of gravitational-wave and electromagnetic observations of white dwarf binaries}

\author[P. T. Johnson et al.]{Peyton T. Johnson,$^{1}$\thanks{E-mail: joh15016@umn.edu}
Michael W. Coughlin,$^{1}$
Ashlie Hamilton,$^{1}$
\newauthor
Mar\'ia Jos\'e Bustamante-Rosell,$^{2}$
Gregory Ashton,$^{3}$
Samuel Corey,$^{1}$
\newauthor
Thomas Kupfer,$^{4}$
Tyson B. Littenberg,$^{5}$
Draco Reed,$^{1}$
Aaron Zimmerman$^{2}$
\\
$^{1}$School of Physics and Astronomy, University of Minnesota, Minneapolis, Minnesota 55455, USA\\
$^{2}$Center for Gravitational Physics, University of Texas at Austin, 2515 Speedway, C1600, Austin, TX 78712, USA\\
$^{3}$Department of Physics, Royal Holloway, University of London, TW20 0EX, United Kingdom\\
$^{4}$Department of Physics and Astronomy, Texas Tech University, PO Box 41051, Lubbock, TX 79409, USA\\
$^{5}$NASA Marshall Space Flight Center, Huntsville, Alabama 35811, USA\\
}

\date{Accepted XXX. Received YYY; in original form ZZZ}

\pubyear{2022}

\begin{document}
\label{firstpage}
\pagerange{\pageref{firstpage}--\pageref{lastpage}}
\maketitle

\begin{abstract}
The upcoming \textit{Laser Interferometer Space Antenna} (\textit{LISA}) will detect a large gravitational-wave foreground of Galactic white dwarf binaries. These sources are exceptional for their probable detection at electromagnetic wavelengths, some long before \textit{LISA} flies. Studies in both gravitational and electromagnetic waves will yield strong constraints on system parameters not achievable through measurements of one messenger alone. In this work, we present a Bayesian inference pipeline and simulation suite in which we study potential constraints on binaries in a variety of configurations. We show how using \textit{LISA} detections and parameter estimation can significantly improve constraints on system parameters when used as a prior for the electromagnetic analyses. We also provide rules of thumb for how current measurements will benefit from \textit{LISA} measurements in the future.
\end{abstract}


\section{Introduction}
The upcoming \textit{Laser Interferometer Space Antenna} (\textit{LISA}) will revolutionize gravitational-wave astronomy \citep{AmAu2017}, opening a completely new frequency band beyond what has been studied so far using ground-based gravitational-wave interferometers such as Advanced LIGO \citep{aLIGO} and Advanced Virgo \citep{aVirgo} and pulsar timing arrays such as NANOGrav \citep{ArBa2020} and the Parkes Pulsar Timing Array \citep{GoSh2021}. Planned for a nominal mission of $4$ years and an extended mission $6$ years further, \textit{LISA} will detect not only extreme mass ratio inspirals or supermassive black hole mergers at low masses \citep{RuBe2010,Marsh2011,NiVa2012,Ses2021}, but also compact objects covering $\sim$ $0.1$ Hz to $10$ mHz, the dominant source class of which is double white dwarf binaries, which have two stellar-mass compact objects with orbital periods less than $1$ hour. Their relatively short orbital periods slowly undergo orbital decay due to the emission of gravitational radiation. This is in addition to binaries that contain other compact objects such as hot sub-dwarf stars, neutron stars and possibly black holes. Short period binaries whose parameters we can measure in advance are known as ``verification'' sources \citep{KuKo2018}, as their gravitational-wave strain can be predicted based on parameters constrained by electromagnetic observations.

Time-domain, optical surveys such as the Asteroid Terrestrial$-$impact Last Alert System (ATLAS, \citealt{ToDe2018}) and the Zwicky Transient Facility (ZTF; \citealt{BeKu2018,GrKu2019,MaLa2018}), among others, are detecting white dwarf binaries regularly, with more than a dozen sources already known \citep{BuCo2019,BuFu2019,CoBu2020,BuPr2020}. There are three types of systems that these surveys detect: eclipsing detached systems,  ellipsoidal detached systems, and accreting systems (e.g., AM CVn systems, where a white dwarf accretes hydrogen-poor matter from a compact companion star). After their detection by the survey, these systems are followed-up with high cadence photometry by using instruments such as the Kitt Peak EMCCD Demonstrator \citep{CoDe2019} and CHIMERA \citep{HaHa2016} to measure their orbital decay through measurement of changes in the orbital phase of the binary. These measurements, typically focusing either on the ``eclipse'' times for eclipsing systems or fits to the sinusoidal modulation phase, are then used to construct observed-minus-computed diagrams where the deviation in phase is measured relative to an object without orbital changes.

As white dwarf binaries will form a gravitational-wave foreground, understanding their contribution to the \textit{LISA} spectrum will be important so they can be removed, studying the fainter and rarer signals underneath \citep{LiCo2020}. Population synthesis results point to more than ${\sim} 10,\!000$ binaries expected to be individually resolvable, with all binaries with periods shorter than $15$ minutes expected to be detected, no matter its location in our Galaxy \citep{LaBl2019}. Studying the population of white dwarf binaries is also interesting astrophysically. As inherently quantum objects, they probe quantum mechanics in a regime difficult to replicate on Earth due to the very high temperatures and densities involved, e.g. \citep{ChHw2020}. They are also likely to be the progenitors of type Ia supernovae \citep{Shen2015}, although the exact channel remains uncertain. In addition, they probe white dwarf structure \citep{FuLa2011}, galactic structure \citep{BrMi2019}, binary stellar evolution \citep{NeTo2005,KrCh2018,Ban2017,AnTo2017}, accretion physics \citep{CaNe2015} and general relativity \citep{BuCo2019,KuBa2019}.

In general, \textit{LISA} will make it possible to identify many white dwarf binaries that were bright enough to be picked up by optical surveys, but only through measurements of sky location, period, distance, and frequency will they be able to be identified as such in the surveys. Many previous works have pointed out that electromagnetic and gravitational-wave measurements will provide complementary views of the white dwarf binary population, e.g. \citep{ShSl2012,ShNe2014}. For example, \citep{ShNe2014} have used Fisher-matrix based analyses to show improvements on parameter uncertainties accessible to both detectors, including distance to the source and masses of the objects. However, most previous analyses using Fisher matrices are limited by the technique, as they only hold in the limit of strong signals with Gaussian noise, and may underestimate the parameter uncertainties.

In this work, we will extend work of this type by including a Bayesian inference based analysis based on state of the art data analysis pipelines built for \textit{LISA}. In Sec.~\ref{sec:pipeline}, we describe the simulation and Bayesian inference pipeline we use for this study. In Sec.~\ref{sec:results}, we discuss the results of the analysis and implications for future observations with \textit{LISA}. Sec.~\ref{sec:conclusion} summarizes our conclusions and forward outlook.

\section{Simulation and Data Analysis Pipeline}
\label{sec:pipeline}
\begin{figure}
    \includegraphics[width=8cm]{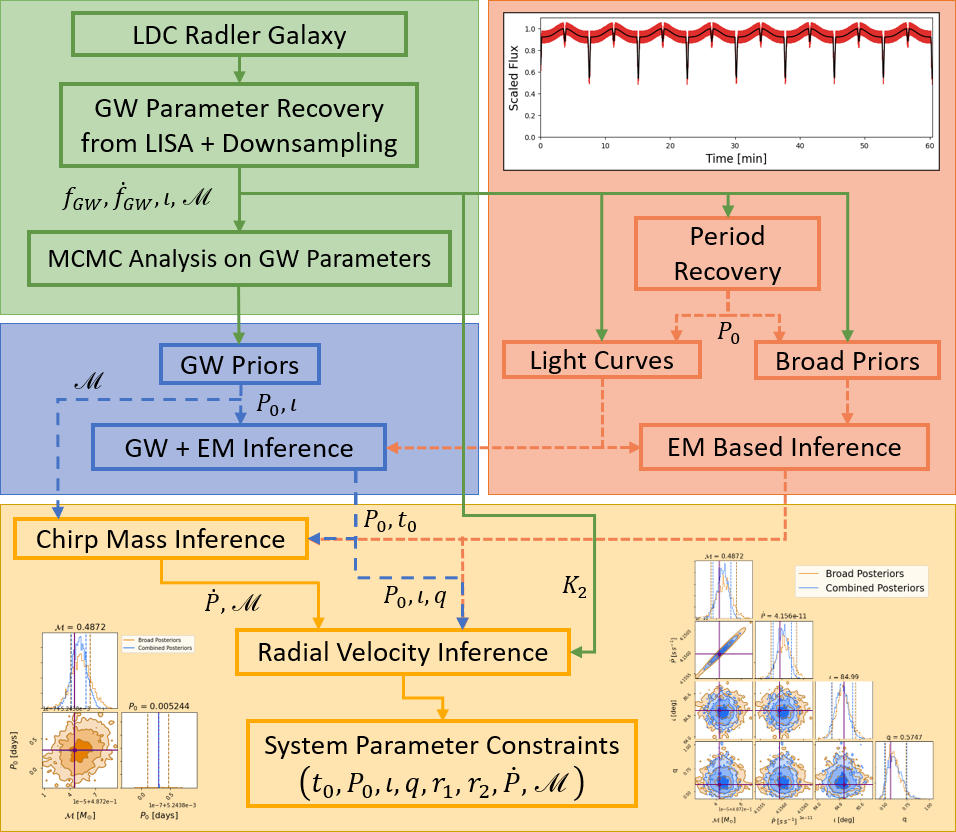}
    \caption{Flow chart of the data analysis pipeline. The pipeline starts with the LDC Radler Galaxy and each of the other steps in the top left panel involves only the gravitational-wave portion of the analysis. In the top right and center left panels are steps pertaining to purely electromagnetic analysis and to joint gravitational-wave and electromagnetic analysis, respectively. The bottom panel displays the Bayesian inferences that are applied to the results of both the purely electromagnetic and joint analysis.}
    \label{fig:flowchart}
\end{figure}

Our work centers around a data analysis pipeline as illustrated in Fig.~\ref{fig:flowchart}.

\subsection{Gravitational-wave analysis}
For our analysis, we use the \texttt{gbfisher} module from \texttt{ldasoft} to simulate ${\sim}20,\!000$ white dwarf binary systems, with all of the parameters described below. Due to computational limitations, we narrowed down our multi-messenger analysis to a subsample consisting of ${\sim} 200$ binaries. For demonstration purposes, we construct two sets of binaries, an eclipsing set and a non-eclipsing set. White dwarf binaries with inclinations between $0^\circ$ and $80^\circ$ are placed in the non-eclipsing set while binaries with inclinations between $80^\circ$ and $90^\circ$ are placed in the eclipsing set. Both sets are narrowed down further by removing all binaries with periods less than $4$ minutes or greater than $20$ minutes as well as binaries with an SNR of less than $20$ or greater than $150$. A subset of $100$ binaries is randomly sampled from each of the eclipsing and non-eclipsing sets. We note that these sets are not meant to be representative of the true population; future analyses will focus on such representative sets.

We use the \texttt{gbmcmc} module from \texttt{ldasoft} to provide the gravitational-wave parameter estimates for our simulated set of white dwarf binaries. \texttt{gbmcmc} uses Reversible Jump Markov Chain Monte Carlo (MCMC) to identify the range of plausible models for each binary in the downsampled set. Next, \texttt{gbmcmc} provides posterior distributions for the recovered parameters which include the initial gravitational-wave frequency, $f_{GW}$, the time derivative of frequency, $\dot{f}_{GW}$, the gravitational-wave amplitude, $A$, the inclination, $\iota$, the polarization angle, the initial gravitational-wave phase, and the ecliptic colatitude and longitude. From these quantities, we can derive the chirp mass, $\mathcal{M}$, which is related to the parameters $f_{GW}$ and $\dot{f}_{GW}$ by
\begin{equation}\label{eq:ChirpMass}
    \mathcal{M} = \frac{c^3}{G} \left(\frac{5}{96} \pi^{-\frac{8}{3}} f_{GW}^{-\frac{11}{3}} \dot{f}_{GW}\right)^{\frac{3}{5}}.
\end{equation}
In addition, we can use these quantities to derive the time rate of change of period given by
\begin{equation}\label{eq:Pdot}
    \dot{P}_0 = \frac{2 \dot{f}_{GW}}{f_{GW}^2} = \left(\frac{\dot{f}_{GW}}{f_{GW}}\right) P_0,
\end{equation}
which will be useful for comparison with the $\dot{P}_0$ and $\mathcal{M}$ parameters recovered by optical surveys. 

\subsection{Light curve analysis}
We produce simulated light curves for the white dwarf binaries using the python package \texttt{ellc} \citep{Max2016}. The light curve generation depends primarily on the initial period $P_0$, the mid-eclipse time of the primary eclipse, $t_0$, the inclination, $\iota$, the mass ratio, $q$, and the ratios of the radii of the primary and secondary to the semi-major axis, $r_1$ and $r_2$, respectively. Additional parameters in the light curve model include the surface brightness ratio $J$, the limb darkening coefficients $\textrm{ldc}_1$ and $\textrm{ldc}_2$, the gravity darkening coefficients, $\textrm{gdc}_1$ and $\textrm{gdc}_2$, and coefficients for the simplified reflection model, $\textrm{heat}_1$ and $\textrm{heat}_2$.

For the analysis of simulated light curves for our objects, we simulate two different sets of observations. The first is a long-baseline cadence simulated over roughly $8$ years with an average observational cadence of $3$ days and noise consistent with expected Zwicky Transient Facility $g$-band measurements. This simulates the type of photometric data expected for using optical surveys such as the Zwicky Transient Facility or the Vera Rubin Observatory for identification of white dwarf binaries. This long baseline data is also most useful for identifying the object's period, $P_0$. We note that here we correct the orbital period evolution using a Post-Newtonian approximation to find that the period evolves according to
\begin{equation}\label{eq:PostPeriod}
    P = P_0 \left(1 - \frac{t}{\tau}\right)^{\frac{3}{8}},
\end{equation}
where $\tau$ is an approximation of the gravitational-wave inspiral timescale given by
\begin{equation*}
    \tau = -\frac{3}{8} \frac{P_0}{\dot{P}}.
\end{equation*}

The second type of simulated observation is high-cadence follow-up data such as provided by KPED \citep{CoDe2019} and CHIMERA \citep{HaHa2016}. For each binary, we construct $25$ sets of simulated observations taken over $1$ night of observations, on average captured at intervals of $120 \pm 5$ days. 

Amongst other parameters, these observations capture the mid-eclipse times $t_i$, for each set of nightly data. These $t_i$ estimates are useful for estimating $\dot{P}_0$ and therefore $\dot{f}_{GW}$; the estimates can be related to the mid-eclipse time observations by
\begin{multline}\label{eq:Residual}
    \Delta t_{eclipse}\left(t_i - t_0\right) = \left(\frac{1}{2} \dot{f}_{GW}\!\!\left(t_0\right) \left(t_i - t_0\right)^2 \right. \\
    \left. + \frac{1}{6} \ddot{f}_{GW}\!\!\left(t_0\right) \left(t_i - t_0\right)^3 + \ldots\right) P\left(t_0\right).
\end{multline}
where $P\left(t_0\right)$ is the orbital period at the reference epoch, $f_{GW}\!\!\left(t_0\right)$, $\dot{f}_{GW}\!\!\left(t_0\right)$, etc, are the orbital frequency and its time derivatives at the reference epoch, and $t_i - t_0$ is the time elapsed since the reference epoch. We note that both the simulated survey and high cadence observations account for the change in period.

The high-cadence photometry also provides constraints on the orbital inclination. While the gravitational-wave recoveries are sensitive to binary orientation such that they range from $0^\circ$ to $180^\circ$, the light curves are not capable of distinguishing between a system facing towards and a system facing away from an observer. For this reason, when using our gravitational-wave observations as priors for the electromagnetic analysis, we map the gravitational-wave inclination posteriors onto the interval $0^\circ$ through $90^\circ$ using the rescaling $\iota^\prime = 90^\circ - \left\lvert\, \iota - 90^\circ \,\right\rvert$ where $\iota^\prime$ is the gravitational-wave inclination, $\iota$, mapped onto the $0^\circ$ to $90^\circ$ interval.

The light curves are sensitive to a number of parameters that the gravitational-wave observations are not. For example, in systems undergoing strong ellipsoidal deformation, the light curve observations loosely constrain the binary's mass ratio, $q = \frac{m_2}{m_1}$. For this reason, we draw $q$ randomly for each binary from a uniform distribution extending from $0.25$ to $1$. Occasionally, this method would yield a mass ratio which causes the primary mass to exceed the Chandrasekhar limit; in these cases, we increase the lower bound for the mass ratio such that the Chandrasekhar limit can't be exceeded.

The light curves are also sensitive to the scaled radii $r_1$ and $r_2$ of the system. To derive these values, we use the system's chirp mass and simulated mass ratio to calculate the individual masses, $m_1$ and $m_2$, from the expressions
\begin{align}
    m_1 &= \left(1+q\right)^{\frac{1}{5}} q^{-\frac{3}{5}} \mathcal{M} \label{eq:Mass1} \\
    m_2 &= \left(1+q\right)^{\frac{1}{5}} q^{\frac{2}{5}} \mathcal{M} \label{eq:Mass2}
\end{align}
We obtain estimates for the radii of each white dwarf, $R_1$ and $R_2$, by fitting a univariate spline curve to a set of white dwarf masses and their corresponding radii and then evaluating the spline for masses $m_1$ and $m_2$, respectively. We obtain a rough approximation of the semi-major axis $a$ by using the fact that the GW frequency is twice the orbital frequency and solving Kepler's Third law to get the expression:
\begin{equation}\label{eq:SMA}
    a = \left[\frac{G \left(m1+m2\right)}{\left(\pi f_{GW}\right)^2}\right]^{\frac{1}{3}}.
\end{equation}
We then scale the white dwarf radii in terms of the semi-major axis to acquire the dimensionless scaled radii values given by $r_1 = \frac{R_1}{a}$ and $r_2 = \frac{R_2}{a}$.

The light curves produced by \texttt{ellc} provide an estimate for flux as a function of time. To provide realistic error bars for the analyses, we take CHIMERA data collected for the $6.9$ minute binary \citep{BuCo2019} from July 2018 and superimpose those error bars upon the simulated light curve. When performing the inference, we also include an arbitrary scaling parameter to account for any offsets in the flux due to the way the photometry is compared to the neighbor star.

\subsection{Combined gravitational-wave and electromagnetic analysis}
We employ the python package \texttt{bilby} \citep{AsHu2019} to perform Bayesian inference on the simulated white dwarf binary light curves. For our analysis, we analyse the light curves using two sets of priors: the first is a broad set of priors designed to be uninformative, and the second uses the gravitational-wave posteriors obtained from \texttt{gbmcmc} as priors for the electromagnetic analysis. The former case simulates the situation we are currently in, where the gravitational-wave data is unavailable or the white dwarf binary is not detected in gravitational-waves, while the latter simulates the utilization of both gravitational-wave and electromagnetic data to improve binary system parameter estimates.

In the case of the broad priors, we use a distribution which is uniform in cosine of inclination from $0$ to $1$ as our inclination prior. We use the python package \texttt{periodfind} \citep{CoBu2021}, a GPU-based implementation of the variance analysis of variance (AOV, \citealt{ScCz1998}) algorithm, to estimate the period, $P_0$, and its uncertainty, $\sigma_{P_0}$ for a particular object. Using these results, we then construct a broad Gaussian period prior with a mean of $P_0$ and a standard deviation of $\sigma_{P_0}$. In the case of the gravitational-wave based priors, we perform a Gaussian kernel density estimate of both the inclination and period posteriors from \texttt{gbmcmc} and we use these to construct the parameter distributions used as our inclination and period priors. In both analyses, each of the remaining parameters, mid-eclipse time, mass ratio, radii, and the scale factor have uniform priors. The uniform prior for the mid-eclipse time extends from $t_0 - \frac{P_0}{2}$ to $t_0 + \frac{P_0}{2}$, for the mass ratio the prior extends from $0.15$ to $1$, and for both the scaled radii and the scale parameter, the priors extend from $0$ to $1$.

In order to widen the overall parameter space and keep our sampling as unbiased as possible, the remaining parameters are randomly generated due to the difficulties in constructing model-based surface brightness ratios, limb-darkening and gravity-darkening coefficients, and reflection coefficients for the light curves. For each individual binary, the surface brightness ratio, the limb-darkening coefficients, and the gravity-darkening coefficients are randomly generated from the range between $0$ and $1$ and for the light curve analyses, uniform priors extending from $0$ to $1$ are used for each of these parameters. Similarly, the reflection model coefficients are randomly generated from the range between $0$ and $5$ with uniform priors from $0$ to $5$ used for the light curve analyses.

A Gaussian likelihood function is used, appropriate for the error bars associated with optical data, computed by comparing the flux and flux uncertainties to the simulated light curve model. We vary the parameters $P_0$, $t_0$, $\iota$, $q$, $r_1$, $r_2$, $J$, $\textrm{ldc}_1$, $\textrm{ldc}_2$, $\textrm{gdc}_1$, $\textrm{gdc}_2$, $\textrm{heat}_1$, $\textrm{heat}_2$, and the scale factor during the inference. To carry out the Bayesian inferences we use the python package \texttt{bilby}, which uses the python package \texttt{PyMultiNest} \citep{BuGe2014} based on the C-library \texttt{MultiNest} \citep{FeHo2009} as its backend, shown to be useful for high-dimensional sampling problems in many areas of astrophysics.

\subsection{Combining multi-night observations}
Using the expression for chirp mass given by equation~\ref{eq:ChirpMass} and $\Delta t_{eclipse}$ given by equation~\ref{eq:Residual}, we use the residual eclipse times derived from each observation to fit for the chirp mass and initial period of each white dwarf binary system. To do so, we construct a Gaussian likelihood using the median and standard deviation of the eclipse time residuals calculated from each observation. For our initial period priors we construct Gaussian kernel density estimates of the period posteriors obtained from the light curve fitting process. For the electromagnetic analyses we used a uniform prior for chirp mass extending from $0.05$ to $1.25$ solar masses, for the combined analyses we construct a Gaussian kernel density estimate of the chirp mass distributions constructed using the $f_{GW}$ and $\dot{f}_{GW}$ posteriors obtained from \texttt{gbmcmc}. 

\subsection{Radial velocities}
The final set of simulated observations are radial velocity observations of the white dwarf binaries. In general, these are required to make accurate estimates of the individual masses of the system. The radial velocity of the secondary object, $K_2$, is related to the orbital period, chirp mass, mass ratio, and inclination by
\begin{equation}\label{eq:RadialVelocity}
    K_2 = \left(\frac{2 \pi G \mathcal{M}}{P_0}\right)^{\frac{1}{3}} \frac{\sin\iota}{q^{\frac{1}{5}} \left(1+q\right)^{\frac{3}{5}}}.
\end{equation}
Passing the system's chirp mass, the simulated mass ratio, the inclination recovered by \texttt{gbfisher}, and the period recovered by \texttt{gbfisher} into equation~\ref{eq:RadialVelocity} gives us the radial velocity of the secondary white dwarf. We construct a Gaussian likelihood using the period recovered by \texttt{periodfind} as our input data and a fixed estimate of uncertainty on the radial velocity of $\pm 50 \textrm{ km s}^{-1}$. We construct Gaussian kernel density estimates of the inclination and mass ratio posterior obtained from the light curve fitting process. Additionally, we construct a Gaussian kernel density estimate of the chirp mass posterior obtained from fitting the residual eclipse times. Then we use the kernel density estimates to produce chirp mass, inclination, and mass ratio priors which are used along with the likelihood to carry out a Bayesian inference.

\section{Results}
\label{sec:results}
For analysis and interpretation purposes, the white dwarf binary systems can largely be categorized as either eclipsing or non-eclipsing systems. Therefore, in the following, we will generally separate out our conclusions for each object type for the different parameters.

\newcommand{\+}{+\:\!}
\begin{table}
    \centering
    \renewcommand{\arraystretch}{2}
    \setlength{\tabcolsep}{0.4em}
    \begin{tabular}{| c c c c |}
        \hline
        ~ & Gravitational-Wave & Electromagnetic & Combined \\
        \hline
        \Centerstack{$\iota$ \\ [$^\circ$]} & 
        $82.34 \substack{\+ 2.63 \\ -2.89}$ & 
        $84.29 \substack{\+ 1.81 \\ -1.02}$ & 
        $83.96 \substack{\+ 0.86 \\ -0.75}$ \\ 
        
        \Centerstack{$P_0$ \\ [s]} & 
        $691.77277 \substack{\+ 0.00017 \\ -0.00017}$ & 
        $691.77270510 \substack{\+ 0.00000064 \\ -0.00000062}$ & 
        $691.77277 \substack{\+ 0.00017 \\ -0.00018}$ \\ 
        
        \Centerstack{$\dot{P}$ \\ [$\textrm{s\,s}^{-1}$]} & 
        $0.51 \substack{\+ 0.14 \\ -0.14} \times 10^{-11}$ & 
        $0.45151 \substack{\+ 0.00060 \\ -0.00061} \times 10^{-11}$ & 
        $0.45148 \substack{\+ 0.00054 \\ -0.00059} \times 10^{-11}$ \\ 
        
        \Centerstack{$\mathcal{M}$ \\ [$M_{\odot}$]} & 
        $0.211 \substack{\+ 0.032 \\ -0.036}$ & 
        $0.19640 \substack{\+ 0.00016 \\ -0.00016}$ & 
        $0.19639 \substack{\+ 0.00014 \\ -0.00015}$ \\ 
        
        $q$ & 
        $~$ & 
        $0.83 \substack{\+ 0.16 \\ -0.22}$ & 
        $0.82 \substack{\+ 0.18 \\ -0.22}$ \\ 
        
        \Centerstack{$R_1$ \\ [$R_{\odot}$]} & 
        $~$ & 
        $1.91 \substack{\+ 0.33 \\ -0.46} \times 10^{-2}$ & 
        $1.83 \substack{\+ 0.32 \\ -0.33} \times 10^{-2}$ \\ 
        
        \Centerstack{$R_2$ \\ [$R_{\odot}$]} & 
        $~$ & 
        $1.99 \substack{\+ 0.35 \\ -0.41} \times 10^{-2}$ & 
        $2.09 \substack{\+ 0.26 \\ -0.30} \times 10^{-2}$ \\ 
        \hline
    \end{tabular}
    \caption{Table of parameters for the example eclipsing white dwarf binary shown in Fig.~\ref{fig:eclipsing_plots}.}
    \label{tab:eclipsing}
\end{table}
\begin{table}
    \centering
    \renewcommand{\arraystretch}{2}
    \setlength{\tabcolsep}{0.3em}
    \begin{tabular}{| c c c c |}
        \hline
        ~ & Gravitational-Wave & Electromagnetic & Combined \\
        \hline
        \Centerstack{$\iota$ \\ [$^\circ$]} & 
        $33.9 \substack{\+ 17.1 \\ -23.2}$ & 
        $13.9 \substack{\+ 10.7 \\ -8.6}$ & 
        $13.2 \substack{\+ 9.7 \\ -7.6}$ \\ 
        
        \Centerstack{$P_0$ \\ [s]} & 
        $650.24647 \substack{\+ 0.00011 \\ -0.00011}$ & 
        $650.24639894 \substack{\+ 0.00000052 \\ -0.00000055}$ & 
        $650.24646 \substack{\+ 0.00011 \\ -0.00011}$ \\ 
        
        \Centerstack{$\dot{P}$ \\ [$\textrm{s\,s}^{-1}$]} & 
        $0.687 \substack{\+ 0.086 \\ -0.085} \times 10^{-11}$ & 
        $0.62 \substack{\+ 0.26 \\ -0.28} \times 10^{-11}$ & 
        $0.680 \substack{\+ 0.082 \\ -0.084} \times 10^{-11}$ \\ 
        
        \Centerstack{$\mathcal{M}$ \\ [$M_{\odot}$]} & 
        $0.237 \substack{\+ 0.018 \\ -0.018}$ & 
        $0.222 \substack{\+ 0.053 \\ -0.066}$ & 
        $0.236 \substack{\+ 0.017 \\ -0.018}$ \\ 
        
        $q$ & 
        $~$ & 
        $0.67 \substack{\+ 0.32 \\ -0.42}$ & 
        $0.65 \substack{\+ 0.32 \\ -0.40}$ \\ 
        
        \Centerstack{$R_1$ \\ [$R_{\odot}$]} & 
        $~$ & 
        $1.07 \substack{\+ 1.99 \\ -0.99} \times 10^{-2}$ & 
        $1.19 \substack{\+ 2.28 \\ -1.06} \times 10^{-2}$ \\ 
        
        \Centerstack{$R_2$ \\ [$R_{\odot}$]} & 
        $~$ & 
        $0.60 \substack{\+ 1.54 \\ -0.53} \times 10^{-2}$ & 
        $0.77 \substack{\+ 1.65 \\ -0.66} \times 10^{-2}$ \\
        \hline
    \end{tabular}
    \caption{Table of parameters for the example non-eclipsing white dwarf binary shown in Fig.~\ref{fig:noneclipsing_plots}.}
    \label{tab:noneclipsing}
\end{table}

\begin{figure}
    \includegraphics[width=7.5cm]{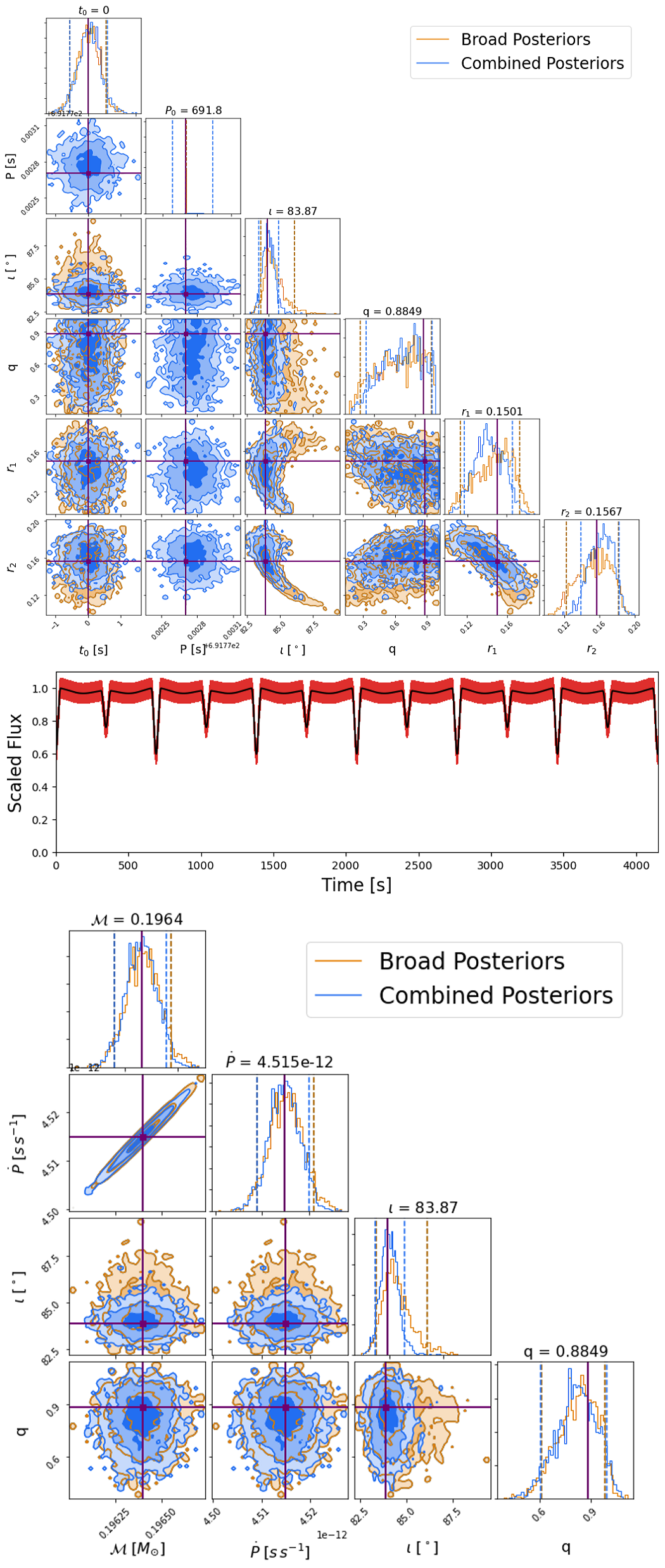}
    \caption{Corner plots and light curve for a sample eclipsing white dwarf binary. Top: Corner plot comparing the posteriors obtained from the light curve inference using the two prior sets. Center: Light curve plot displaying the luminosity as a function of time over several orbits. Bottom: Corner plot comparing the posteriors obtained from the radial velocity inference using the two prior sets.}
    \label{fig:eclipsing_plots}
\end{figure}
\begin{figure}
    \includegraphics[width=7.5cm]{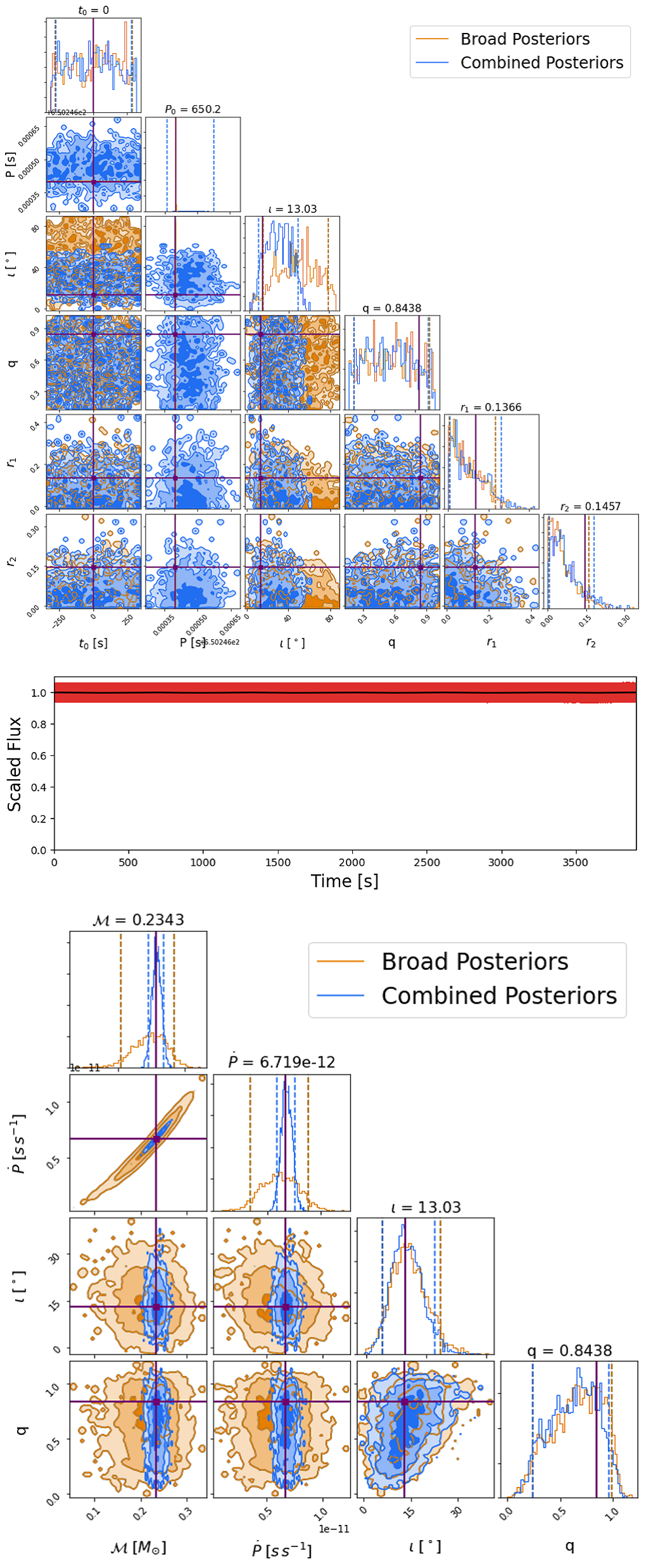}
    \caption{Corner plots and light curve for a sample non-eclipsing white dwarf binary. Top: Corner plot comparing the posteriors obtained from the light curve inference using the two prior sets. Center: Light curve plot displaying the luminosity as a function of time over several orbits. Bottom: Corner plot comparing the posteriors obtained from the radial velocity inference using the two prior sets.}
    \label{fig:noneclipsing_plots}
\end{figure}

\textbf{Period constraints.}
The period constraints are typically several orders of magnitude more precise for the broad and combined posteriors than for the gravitational-wave priors, an effect clearly illustrated in the upper corner plots in Fig.~\ref{fig:eclipsing_plots} and Fig.~\ref{fig:noneclipsing_plots} as well as in Table~\ref{tab:eclipsing} and Table~\ref{tab:noneclipsing}. The uncertainty on the period recovered for the broad and combined posteriors shows that over an $8$ year period we can generally expect to accumulate a total error less than the orbital period; in line with the expectation that orbital cycle count is well established.

\begin{figure*}
    \includegraphics[width=17cm]{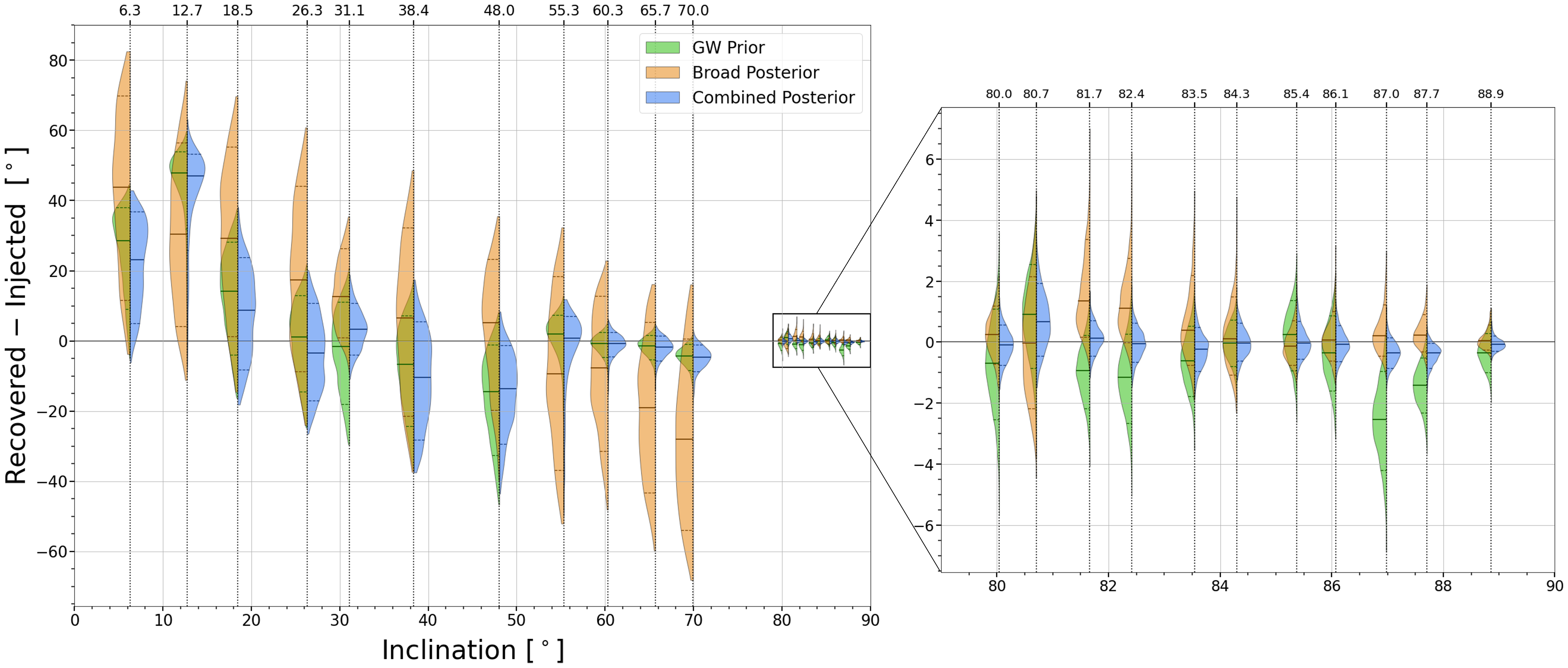}
    \caption{Recovered $-$ injected diagram with the residual inclination distributions from the light curve inference plotted against the system inclinations for a sample of eclipsing and non-eclipsing binaries. Each ``violin'' corresponds to a white dwarf binary system and displays two inclination posteriors corresponding to the two sets of priors as well as the corresponding gravitational-wave inclination prior. The horizontal lines embedded in each distribution mark the $10$th, $50$th, and $90$th percentiles of the distribution.}
    \label{fig:incl_plot}
\end{figure*}

\textbf{Inclination constraints.}
Eclipsing binary light curves, such as the ones magnified on the right in Fig.~\ref{fig:eclipsing_plots}, have strongly constrained inclination, as properties of the light curve such as eclipse duration and eclipse depth are closely related to inclination. The resolvability of the eclipses, especially shallower ones, depends heavily on the noise level of the light curve. In this sense, a system that is ``eclipsing'' depends not only on its angle relative to the detector, but also the detector sensitivity itself. For a handful of the ``eclipsing'' binaries in our simulated set, the recovered parameters are at a precision more akin to the non-eclipsing binaries due to the fact that the eclipses were buried within the noise. Unsurprisingly, Fig.~\ref{fig:incl_plot} shows that the inclinations of the eclipsing binaries, especially those recovered solely through electromagnetic analyses, are much better constrained than for the non-eclipsing binaries. For eclipsing systems with inclinations above ${\sim} 75^\circ$, the inclinations recovered purely from gravitational-wave data and purely from electromagnetic data are both constrained to a precision of within ${\sim} 2^\circ$ of the true inclinations.

For non-eclipsing systems the level of precision on the inclination recovered through gravitational-wave analysis increases with increasing inclination of the systems. For example, the lower inclination binaries in Fig.~\ref{fig:incl_plot} with inclinations below ${\sim} 45^\circ$ display levels of precision on the order of ${\sim} 10^\circ$ whereas non-eclipsing systems with higher inclinations constrain the system inclinations to within a few degrees. In contrast, the constraints on the inclinations recovered from purely electromagnetic analyses of non-eclipsing binaries show little to no correlation with the inclinations of the systems, with precision on the order of ${\sim} 40^\circ$ regardless of inclination.

\begin{figure}
    \includegraphics[width=8cm]{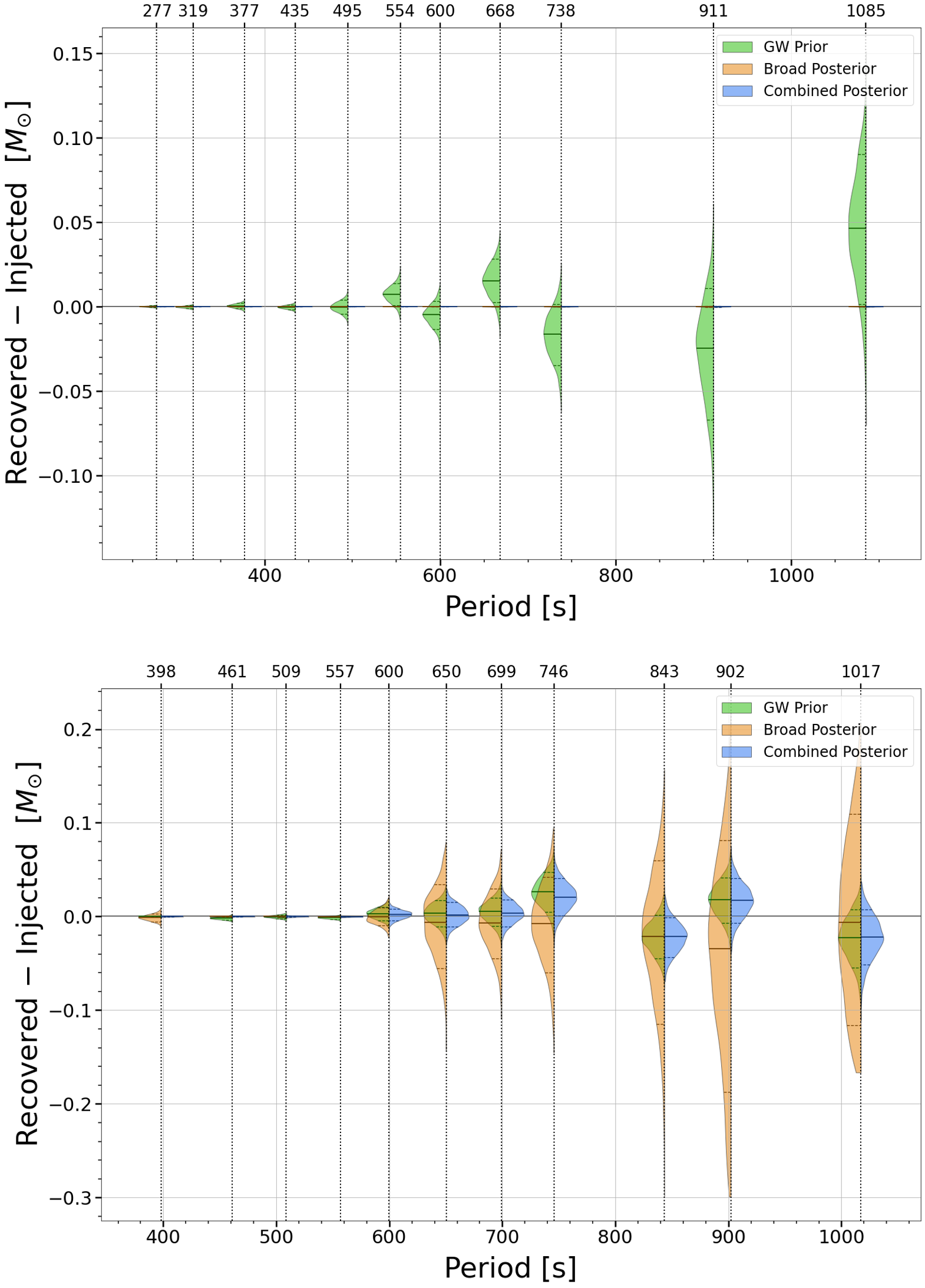}
    \caption{Recovered $-$ injected diagrams with the residual chirp mass distributions plotted against the system orbital periods for a sample of eclipsing (top) and non-eclipsing binaries (bottom). Each ``violin'' corresponds to a white dwarf binary system and displays two chirp mass posteriors corresponding to the two sets of priors as well as the corresponding calculated gravitational-wave chirp mass prior. The horizontal lines embedded in each distribution mark the $10$th, $50$th, and $90$th percentiles of the distribution.}
    \label{fig:chirpmass_plot}
\end{figure}

\textbf{Chirp mass constraints.}
As described above, the potential measurements of $\dot{P}$ yield measurements of chirp mass. Shorter periods tend to lead to better recovery of $\dot{P}$ which in turn leads to more precise measurements of the chirp mass. The trend is displayed prominently in Fig.~\ref{fig:chirpmass_plot} where we show the recovered chirp mass distributions plotted against period; we also see that for eclipsing binaries and short period non-eclipsing binaries, inclusion of optical data reaches and surpasses the level of precision in the chirp mass estimate obtained from the gravitational-wave information alone. In particular, we found that for non-eclipsing binaries with periods below ${\sim} 11$ minutes, the measurement of chirp mass obtained from the high-cadence optical follow-up by itself yielded a level of precision comparable to the measurements obtained from \textit{LISA} analysis alone; an effect which presents itself in the chirp mass of the non-eclipsing binary shown in Fig.~\ref{fig:chirpmass_plot} as well as in Table~\ref{tab:noneclipsing}. In contrast, non-eclipsing binaries with longer orbital periods showed chirp mass recovery precision more akin to the level of precision obtained from gravitational-wave analyses. It is in this regime, where the precision in chirp mass recovery obtained from gravitational-wave and electromagnetic analyses is similar, that the combined analysis proves the most benefit and shows the greatest improvement in chirp mass recovery over the measurements obtained by using each source individually.

\textbf{Mass ratio constraints.}
Using purely gravitational-wave based observations, it is not possible to constrain the mass ratio $q$. Additionally, due to the difficulty of a-priori knowing the gravity-darkening and limb-darkening coefficients, ellipsoidal variations in the light curve offer, at most, a model-dependent constraint on the mass ratio $q$. More promising, however, is the use of mass-radius relations in eclipsing systems with detectable $\dot{P}$, such as for ZTF J2243+5242 \citep{BuCo2020}. Inclusion of a radial velocity, however, allows for small improvements in the mass ratio recovery and thus the uncertainty in the radial velocity dictates to some effect how well the mass ratio is constrained. As expected from equation~\ref{eq:RadialVelocity}, the constraint on the mass ratio improves when the inclination, chirp mass, and period are well constrained. This is the case for most eclipsing binaries, whereas for non-eclipsing binaries, the limited improvements in mass ratio recovery due to the radial velocity constraint are less pronounced.

\section{Conclusion}
\label{sec:conclusion}
We constructed a robust data analysis pipeline designed for use with \textit{LISA} for carrying out joint analyses of gravitational-wave and electromagnetic information from white dwarf binary systems. Using the results of our pipeline, we observed a number of improvements in parameter space estimation offered by using Bayesian inference to carry out a combined analysis. In particular, we saw that combined analyses led to increases in precision in period, inclination, and chirp mass, and quantified these improvements across the parameter space. Additionally, we observed that minor improvements in the constraints on the mass ratio could be made by incorporating radial velocity into the parameter inference, where we took a basic model of radial velocity measurements from potential time-resolved spectroscopy.

While our framework is a strong step forward relative to the current paradigm of using Fisher matrices to make parameter estimates, the pipeline for combining gravitational-wave and electromagnetic analyses is currently limited by the computational power available for running large scale simulations of white dwarf binary populations, as well as the subsequent parameter recovery processes. In the future, we aim to adapt our framework to enable population level studies to bring us closer to the goal of being capable of simulating realistic gravitational-wave data for existing binary systems to make as accurate projections for LISA as possible. Additionally, we will aim to incorporate more sophisticated simulations of spectroscopic data as well as simulated distance estimates such as those provided by Gaia \citep{BrVa2021} into the parameter recovery portion of our pipeline, with the goal of automating the process for the white dwarf binaries that experiments such as ZTF are finding \citep{BuPr2020}.

Looking forward, we intend to build out the light curve analysis used on the simulated data and apply it to non-simulated electromagnetic data collected by systems such as KPED \citep{CoDe2019}. Before \textit{LISA} flies, these observations can be used to track the period evolution and eclipse timing of verification binaries identified now before gravitational-wave data is available. Measurements of this kind will prepare these systems for the first multi-messenger analyses once \textit{LISA} data becomes available. We are using KPED to observe short-period white dwarf binaries in a dedicated program regularly, e.g \citep{BuCo2019,CoBu2020}, and can use these observations to track their period evolution. The period evolution of these objects can be used to look for gravitational-wave emission, or other physical processes that change the period. We look forward to having characterized as many of these systems as possible in preparation for \textit{LISA}.

\section*{Data Availability Statement}
The software described in this article is available at \url{https://github.com/mcoughlin/gwemlisa}. The simulations produced for this article will be shared on reasonable request to the corresponding author.

\section*{Acknowledgements}
Peyton T. Johnson and Ashlie Hamilton thank the Undergraduate Research Opportunities Program at the University of Minnesota for funding their work.
M.~W.~Coughlin acknowledges support from the National Science Foundation with grant numbers PHY-2010970 and OAC-2117997.
Aaron Zimmermann and Mar\'ia Jos\'e Bustamante-Rosell acknowledge support from the National Science Foundation with grant number PHY-1912578.
Portions of this work were performed during the CCA LISA Sprint, supported by the Simons Foundation.
The authors acknowledge the Minnesota Supercomputing Institute\footnote{\url{http://www.msi.umn.edu}} (MSI) at the University of Minnesota for providing resources that contributed to the research results reported within this paper under project ``Identification of Variable Objects in the Zwicky Transient Facility.''

\bibliographystyle{mnras}
\bibliography{references.bib}

\bsp	
\label{lastpage}

\end{document}